\documentclass[%
 amssymb, amsmath,%
 prb,
11pt,  
twocolumn,
 aps,
 unsortedaddress,superscriptaddress,
]{revtex4-1}
\pdfoutput=1

\usepackage[squaren]{SIunits}
\usepackage{graphicx}% Include figure files
\usepackage{dcolumn}% Align table columns on decimal point
\usepackage{bm}% bold math
\usepackage{tabularx}
\usepackage{bm}%
\usepackage{color}      % use if color is used in text
\usepackage[colorlinks=true,linkcolor=blue,citecolor=blue,urlcolor=blue]{hyperref}%
\expandafter\ifx\csname package@font\endcsname\relax\else
 \expandafter\expandafter
 \expandafter\usepackage
 \expandafter\expandafter
 \expandafter{\csname package@font\endcsname}%
\fi
\begin{document}

\title{Single-photon emitting diode in silicon carbide}
\author{A. Lohrmann}
\affiliation{%
School of Physics, The University of Melbourne, Victoria 3010, Australia
}%

\author{N. Iwamoto}
\affiliation{%
SemiConductor Analysis and Radiation Effects Group, Japan Atomic Energy Agency, 1233 Watanuki, Takasaki, Gunma 370-1292, Japan.
}%

\author{Z. Bodrog}
\affiliation{Institute for Solid State Physics and Optics, Wigner Research Centre for Physics, Hungarian Academy of Sciences, Budapest, POB 49, H-1525, Hungary} 

\author{S. Castelletto}
\affiliation{%
School of Aerospace, Mechanical and Manufacturing Engineering RMIT University, Melbourne, Victoria 3001, Australia.
}%

\author{T. Ohshima}
\affiliation{%
SemiConductor Analysis and Radiation Effects Group, Japan Atomic Energy Agency, 1233 Watanuki, Takasaki, Gunma 370-1292, Japan.
}%

\author{T. J. Karle}
\affiliation{%
School of Physics, The University of Melbourne, Victoria 3010, Australia
}%

\author{A. Gali}
\affiliation{Institute for Solid State Physics and Optics, Wigner Research Centre for Physics, Hungarian Academy of Sciences, Budapest, POB 49, H-1525, Hungary} 
\affiliation{%
Department of Atomic Physics, Budapest University of Technology and Economics, Budafoki \'{u}t 8, H-1111, Budapest, Hungary 
}%

\author{S. Prawer}
\affiliation{%
School of Physics, The University of Melbourne, Victoria 3010, Australia
}%

\author{J. C. McCallum}
\affiliation{%
School of Physics, The University of Melbourne, Victoria 3010, Australia
}%

\author{B. C. Johnson}
\affiliation{%
Centre for Quantum Computing and Communication Technology, School of Physics, University of Melbourne, Victoria 3010, Australia.
}%

\begin{abstract}
Electrically driven single-photon emitting devices have immediate applications in quantum cryptography, quantum computation and single-photon metrology. Mature device fabrication protocols and the recent observations of single defect systems with quantum functionalities make silicon carbide (SiC) an ideal material to build such devices. Here, we demonstrate the fabrication of bright single photon emitting diodes. The electrically driven emitters display fully polarized output, superior photon statistics (with a count rate of $>$300~kHz), and stability in both continuous and pulsed modes, all at room temperature. The atomic origin of the single photon source is proposed. These results provide a foundation for the large scale integration of single photon sources into a broad range of applications, such as quantum cryptography or linear optics quantum computing.
\end{abstract}

\maketitle

With breakthroughs in SiC growth technologies and its excellent thermal, mechanical and physical properties\cite{Nakamura2004}, SiC has become an outstanding wide band gap semiconductor for both emerging optical and electronic applications. Well-developed CMOS compatible processing protocols exist for SiC providing the availability of broadband photonic optical cavities\cite{Yamada2011} and, with the possibility of controlled doping of both p- and n-type impurities\cite{Laube2002}, commercialized electronic devices. The compound nature and extensive polytypism of SiC results in a broad array of defects some of which have promising applications in quantum computing, cryptography and metrology. Optically driven single photon emission in 4H-SiC has been demonstrated for the carbon-vacancy carbon-antisite $\text{C}_\text{V}\text{C}_\text{Si}$ pair \cite{Castelletto2013}, the silicon vacancy (V$_\text{Si}$) \cite{Widmann2014, Fuchs2014} and the divacancy ($\rm V_{C}V_{Si}^{0}$) \cite{Christle2014}, and in 3C-SiC for the $\text{C}_\text{V}\text{C}_\text{Si}$ pair\cite{Castelletto2014}. Other single photon emitters in wide band-gap materials such as ZnO\cite{Morfa2012} may also display similar characteristics. \\
% Single photon emission from defects in wide band gap semiconductors have been limited to optical defects in diamond, until recently the discovery of single photon emitters in ZnO\cite{Morfa2012} and SiC\cite{Castelletto2013, Castelletto2014,Widmann2014, Christle2014, Fuchs2014} were reported. 
\begin{figure*}[htbp]
\begin{center}
\includegraphics[width=14cm]{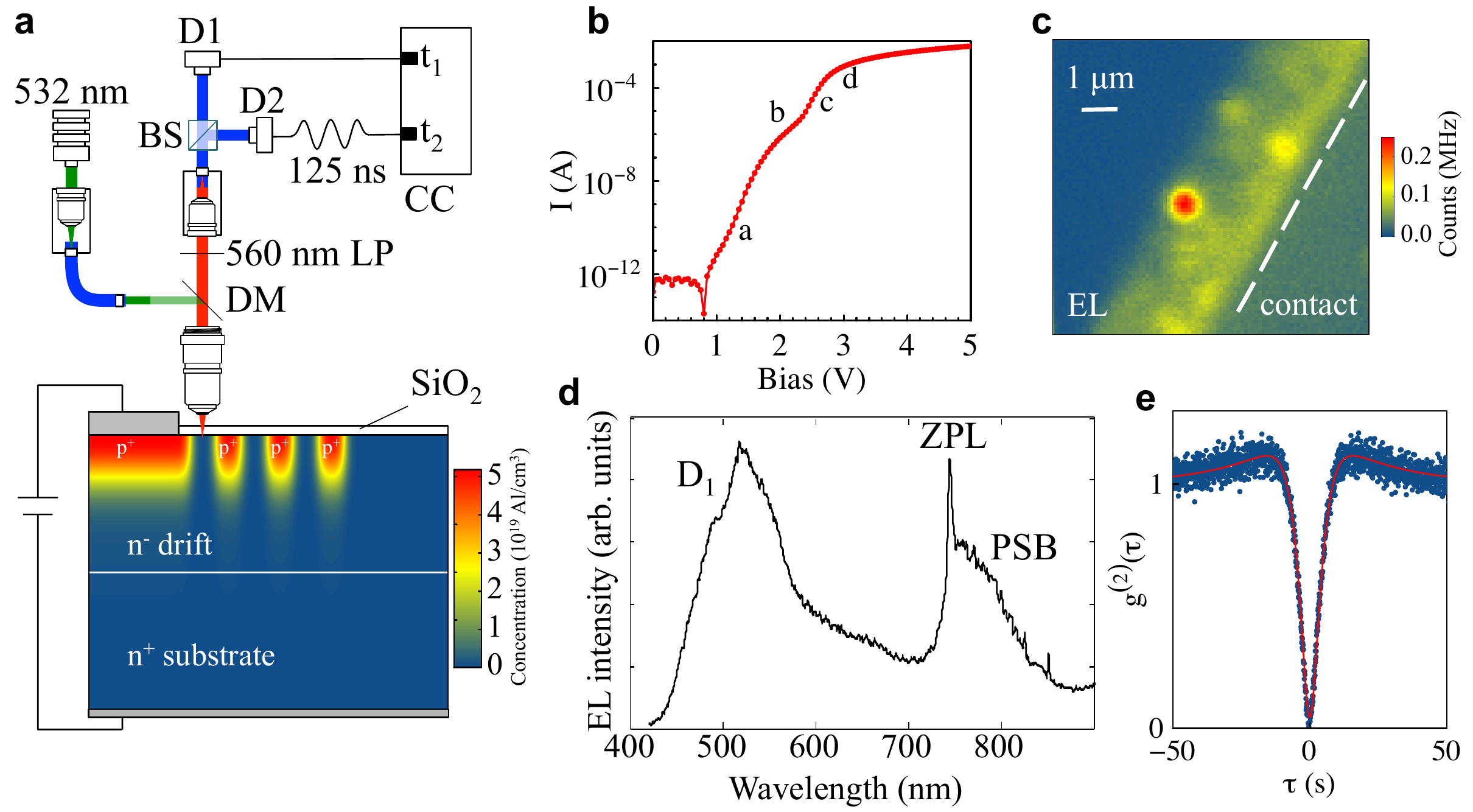}
\end{center}
\caption{ \textbf{Single photon sources in bulk 4H-SiC and fully fabricated p$^{+}$n junction diodes.} \textbf{a}, Schematic of the confocal set-up used to characterize the single photon emitters. It includes a Hanbury, Brown-Twiss interferometer with two single photon avalanche detectors (D1 and D2) connected to a correlation card (CC). The dichroic mirror (DM) was removed when in used in EL mode. A partial schematic of the device consisting of a p$^{+}$ top contact is also shown. Three floating guard rings encircle the central contact to decrease the electric field at the main contacted junction. \textbf{b}, IV-curve of the diode. In this device, features due to shunt resistance, tunnelling current, diffusion current and series resistance are indicated by the letters a-d, respectively. \textbf{c}, EL map of the edge region of a device. \textbf{d}, Room temperature EL spectrum showing the source of the background (D$_1$ line) and the single photon emitter with a ZPL at 745~nm. \textbf{e}, Background corrected anti-bunching traces with $g^{(2)}(\tau=0)<0.1$ indicating excellent single photon emission characteristics.}
\label{Fig1}
\end{figure*}
To significantly enhance the practicality of single photon emission, it is desirable to integrate single photon emitters into a device to allow electrical pumping. Electrically driven quantum dots have been demonstrated\cite{Yuan2002, Heindel2010, Ward2007, Hargart2013, Bennett2005} but these mostly operate at low temperatures. The possibility of electrical excitation at room temperature of the nitrogen vacancy defect in diamond, NV$^0$, has also been shown\cite{Lohrmann2011, Mizuochi2012} but at present there are severe limitations on the types of devices that can be engineered with this material, namely, the inability to create a high active n-type concentration by ion implantation.
In this work, single photon emitters in the visible spectral range are integrated into 4H-SiC p$^{+}$n and n$^{+}$p junction diodes. Results with 6H-SiC appear in the supplementary material. The fabrication of these single photon emitting diodes is achieved by applying well developed CMOS compatible processes including photolithography, ion implantation and annealing. Photon anti-bunching measurements indicate the quantum nature of the single defect. The emitters have high emission rates, are fully polarized and operate with high stability at room temperature, characteristics which are beneficial for quantum cryptography protocols and linear optics quantum computing\cite{Santori2010}.
\newline

\noindent {\bf\large Results} 

\noindent {\bf Single photon source and device integration.} SiC p$^{+}$n junction diodes were formed in 4H SiC n-type epi-layers by implanting aluminum while the substrate was held at 800$^{\circ}$C. To activate the dopants a post-implantation anneal of 1600-1800$^{\circ}$C was employed. During this anneal a carbon capping layer was used to prevent Si sublimation and step bunching formation \cite{Negoro2004}. No irradiation step was performed to create single defects after annealing. Fig.~\ref{Fig1}a shows the device schematic (bottom) together with the confocal microscope used to observe the material's luminescence. The dopant profile is derived from SRIM calculations which are presented in the Supplementary Information. A typical current-voltage (IV) curve from our fabricated diodes is displayed in Fig.~\ref{Fig1}b and shows the expected rectifying behavior of a pn junction diode with a reverse leakage current of around $1\times 10^{-10}\;\rm A/cm^{-2}$. In forward bias, the device displays clear features that are dominated by the series resistance which limits the injected current. Electroluminescence (EL) is also observed when driven above the threshold voltage of $ V \approx$~3.1~V.  Further fabrication and measurement set-up details are provided in the Methods section below.\\
\begin{figure*}
\begin{center}
\includegraphics[width=\textwidth]{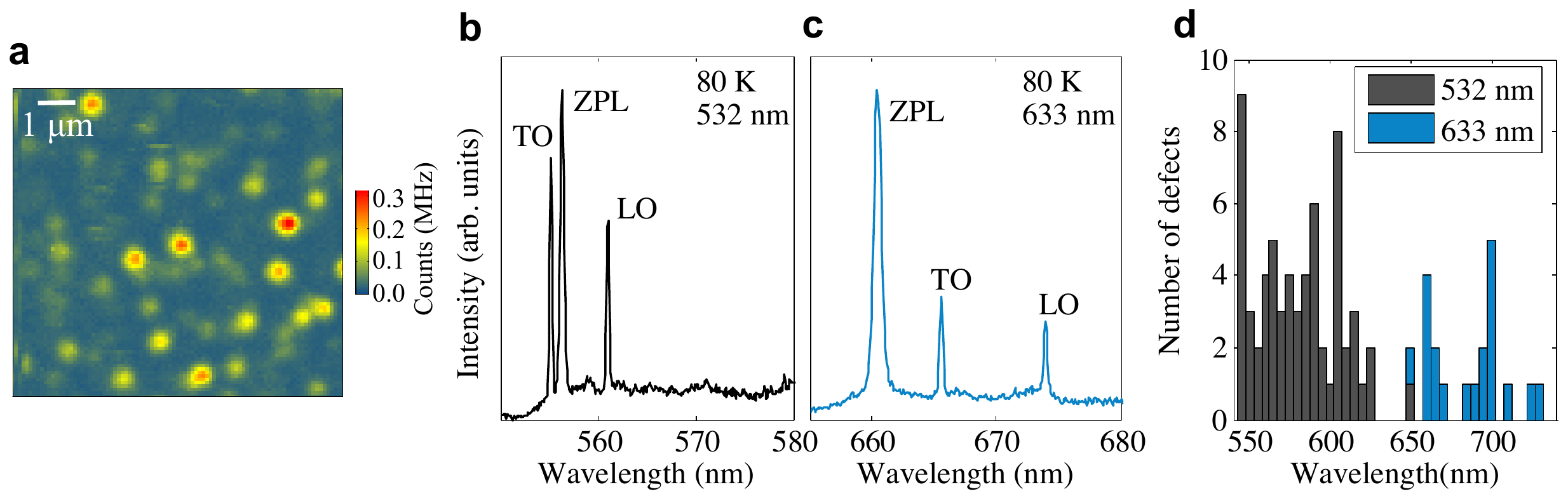}
\end{center}
\caption{{\bf Single defect PL in 4H-SiC at 80~K. a,} Room temperature confocal map excited with 532~nm. {\bf b,c,} Single defect spectra taken at 80~K under excitation with 532~nm and 633~nm for two different defects, respectively. TO and LO indicate Raman lines. The ZPL widths are 0.5~nm for green and 0.8~nm for red excitation. {\bf d}, Histogram created from more than 50 spectra indicating the number of lines per 5~nm interval.} 
\label{Fig2}
\end{figure*}
The EL arises when minority carriers are injected into the n-type region and radiatively recombine as previously observed via deep-level defect states\cite{Fuchs2013}. Fig.~\ref{Fig1}c shows an EL map of the device under an applied forward bias of 10~V. The central bright spot is a single defect with a count rate of 250~kHz and does not display any blinking behaviour. The defects are located near the SiC epilayer surface between the top aluminum contact and the floating guard rings and not in the implanted regions. 

Figure~\ref{Fig1}d shows a typical EL spectrum taken at a single defect site. The spectrum displays two main components. Firstly, a signal arising from the D$_{1}$ center with a weak zero phonon line (ZPL) at 427~nm and a phonon-side band (PSB) with a peak intensity around 520~nm is observed. This often appears as background luminescence in the vicinity of the implanted region and is mostly formed as a result of the implantation induced damage, surviving the high temperature activation anneals. The D$_{1}$ line is associated with a Si antisite (Si$_\text{C}$) \cite{Gali2003}. Secondly, in this example, ZPL emission at a wavelength of 750~nm in Fig.~\ref{Fig1}d from the diffraction limited bright spot appears. To construct the EL map of single photon emitters shown in Fig.~\ref{Fig1}c, emission was collected through a 560~nm long-pass filter to suppress the D$_{1}$ line. Fig.~\ref{Fig1}e shows an example of the background corrected \cite{Beveratos2002} anti-bunching trace with $g^{(2)}(\tau=0)<0.1$ confirming excellent single photon emission characteristics.

While the D$_{1}$ line is relatively well known, a wide variation of ZPL positions and line shapes was observed for single defects (see Supplementary Information). To aid in the identification of these defects we studied their formation in a broad range of SiC materials with different doping densities and processing conditions. For these studies photoluminescence (PL) measurements on high-purity semi-insulating (HPSI) SiC from CREE was performed and enabled a greater number of stable defects to be studied at once. Additionally, when laser excitation was used with the single photon emitting diodes, the single defects could not be clearly characterized as their photo-stability decreased and a significant background luminescence appeared. 

The as-purchased HPSI sample showed only unstable emission as observed previously \cite{Castelletto2013}. As mentioned above, the defects of interest in this work form and are stabilized via the high temperature anneal only. As with the EL emission described in Fig~\ref{Fig1}, over 90\% of these centers are completely photo stable. Further, it was found that if the surface roughness is allowed to degrade during a high temperature anneal without the carbon capping layer \cite{Negoro2004}, the photo-stability and defect density greatly decreased. The defects are thus very sensitive to the state of the surface and detailed confocal maps confirm that they exist in a narrow band close to the surface (see Supplementary Information).

A typical PL confocal map of the HPSI SiC is shown in Fig.~\ref{Fig2}a. The defect density was found to vary slightly for different samples typically in the $2-8\times 10^7\;\rm cm^{-2}$ range. Interestingly, electron irradiation of 2~MeV electrons to a fluence of $1\times 10^{17}\;\rm cm^{-2}$ did not appear to significantly increase the defect density or the type of defects observed after the high temperature annealing.
By lowering the temperature to 80~K, most defect spectra develop sharp ZPL, with line widths ranging from 0.3-2~nm and could be characterized further. Fig.~\ref{Fig2}b and c show two typical single defect spectra taken at 80~K for excitation wavelengths of 532~nm and 633~nm, respectively. Room temperature spectra are presented in the Supplementary Information. Phonon assisted transitions are still present as can be seen by the elevated background signal. The transverse optic (TO) and longitudinal optic (LO) phonon Raman modes are indicated. We performed spectral measurements of more than 50 defects and created a histogram of the recorded ZPL peak positions (Fig.~\ref{Fig2}d). The histogram clearly shows the variety of ZPL over the whole red spectral region. We did not observe any groups of lines as might be expected for different charge states, defects on inequivalent lattice sites or multiple defect types. Instead the lines are continuously distributed over a 200~nm range. The large spread is peculiar and will be addressed in the discussion below. In the following sections we examine several other properties shared by all observed single photon emitters in these materials to give further scope to the discussion.

\begin{figure}
\begin{center}
\includegraphics[width = 0.5\textwidth]{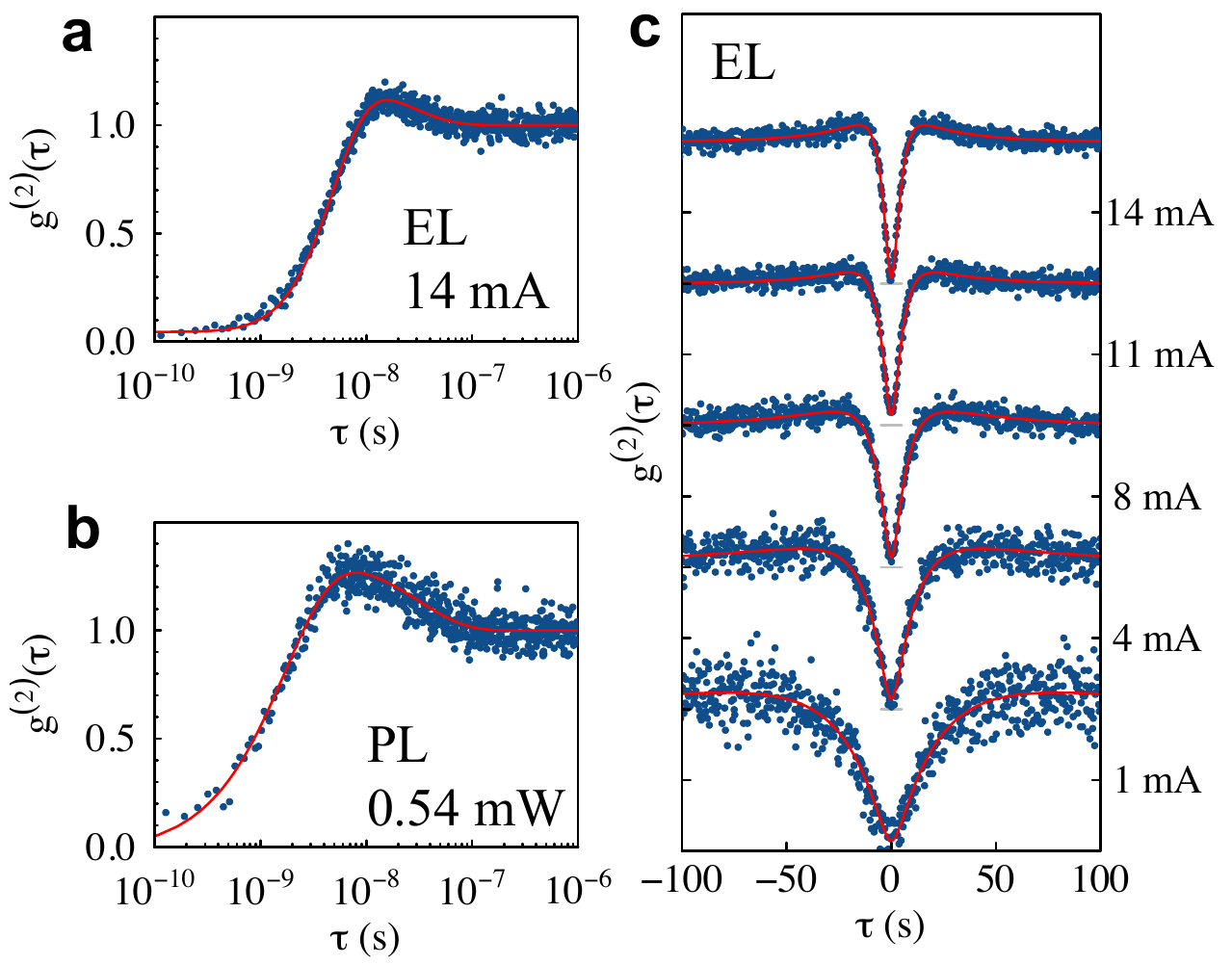}
\end{center}
\caption{ \textbf{Electrically driven single photon emitter dynamics.} Background corrected \textbf{a}, EL and \textbf{b}, PL traces for two spectrally similar single photon emitters in 4H-SiC (solid line fit described in the text). \textbf{c}, EL $g^{(2)}(\tau)$ traces for different currents offset vertically for clarity.}
\label{Fig3}
\end{figure}

\noindent {\bf Single photon emitter dynamics.} Figure~\ref{Fig3}a and \ref{Fig3}b show the background corrected anti-bunching traces of two different single defects in EL (p$^{+}$n device) and PL (HPSI sample) mode that have similar spectral characteristics, respectively. A typical signal-to-background ratio is $\geq$ 0.85. Photon bunching around $\tau=10$~ns is evident in EL and especially in PL modes which is related to a dark metastable state. EL and PL modes differ most strikingly around $\tau=0$ where the EL anti-bunching exhibits a plateau characteristic of a double pump process \cite{Kolesov2012, Mizuochi2012}. Such a process may involve, for example, two step carrier capturing process. The equation that is used to fit the PL data is a double exponential which is commonly used for three level systems \cite{Kitson}. A standard analysis based on such a system yields a relatively fast excited state lifetime of $t_L^{\text{PL}} = 3.3\pm0.3 \text{ ns}$ with transitions to a dark state every 10 cycles. All other centers studied show very similar transition rates ($t_L^{\text{PL}} = 2-5 \text{ ns}$). 

To account for the plateau that is evident in the EL data, a third exponential term is required and yields an excellent fit. Such an empirical relation has a form that may be reflective of a four level electronic system \cite{Kolesov2012, Mizuochi2012}. Fig.~\ref{Fig3}c illustrates the $g^{(2)}(\tau=0)$ evolution under increasing injected current. Due to the faster pump rate, the $g^{(2)}(\tau=0)$ sharpens and there is a subtle increase in the photon bunching. All time constants associated with the fit increase linearly with injected current. 

\begin{figure*}
\begin{center}
\includegraphics[width=14cm]{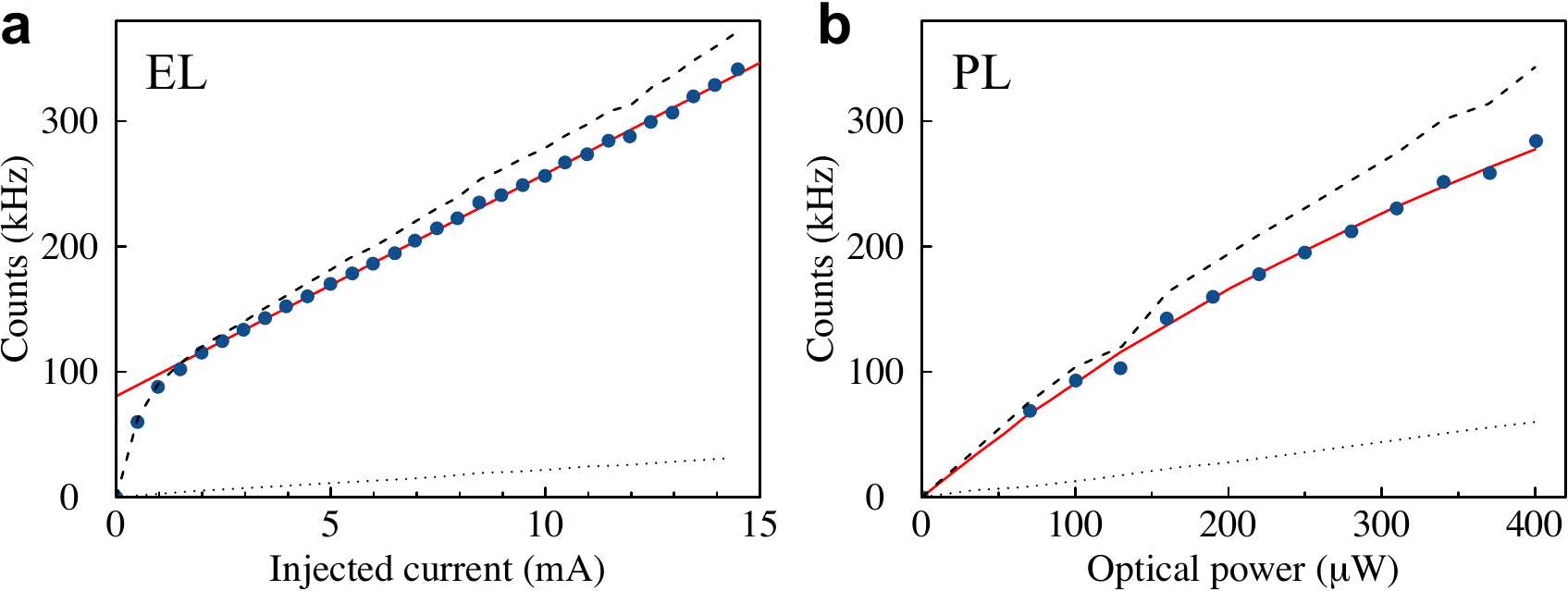}
\end{center}
\caption{{\bf Intensity-excitation power curves for single photon emitters in 4H-SiC.} Intensity curve for a single defect as a function of {\bf a}, current (EL) and {\bf b}, laser power (PL) corrected for the back-ground (dotted line). The raw signal is indicated by the dashed line. }
\label{Fig4}
\end{figure*}

The intensity of the single defect emission in EL varies strongly from defect to defect with many being well over 300~kHz. One contributing factor to the intensity variation is the distance from the p$^{+}$n or n$^{+}$p junction which determines the current density at the defect location. A typical EL intensity-current curve is shown in Fig.~\ref{Fig4}a. After subtracting the linear background that originates from the D$_1$ center emission (dotted line), we obtain a linear relationship. Note, that the initial non-linearity is due to the unknown current density at the defect site and does not indicate a saturating behavior. The detector count rate reached 350~kHz at 14~mA without signs of saturation. This intensity is more than twice as bright as the saturation count rate of an optically driven single nitrogen vacancy (NV$^{-}$) in diamond investigated using the same experimental set-up, and more than 6 times brighter than the EL saturation count rate from NV$^{0}$ defects in diamond\cite{Mizuochi2012}. At higher currents, some degradation in the device characteristics is observed which affects the count rate. Therefore, no true saturation behavior has yet been observed.

In contrast, the intensity-power dependence in PL shown in Fig.~\ref{Fig4}b displays a saturation type behavior over a count rate similar to that observed in EL but with a dependence given by, $C(P) = C_{sat} P / (P+P_{sat})$. Here, $P$ is the excitation power, $C$ the detector count rate, $P_{sat}$ the saturation power and $C_{sat}$ the saturation count rate. Fig.~\ref{Fig4}b shows a typical intensity-power curve with a saturation count rate of $C_{sat} = 0.9\pm0.2$~MHz and saturation power of $P_{sat} = 880\pm 250\;\rm\mu W$. 

Although the defect emission reaches similar count rates in both excitation modes, only the PL mode shows the beginnings of saturation behaviour. The significantly lower branching ratio in EL, indicated by the weak bunching of the $g^{(2)}(\tau)$, can therefore yield an explanation for the significantly higher count rates without saturation. 

\begin{figure}[b]
\begin{center}
\includegraphics[width=8.5cm]{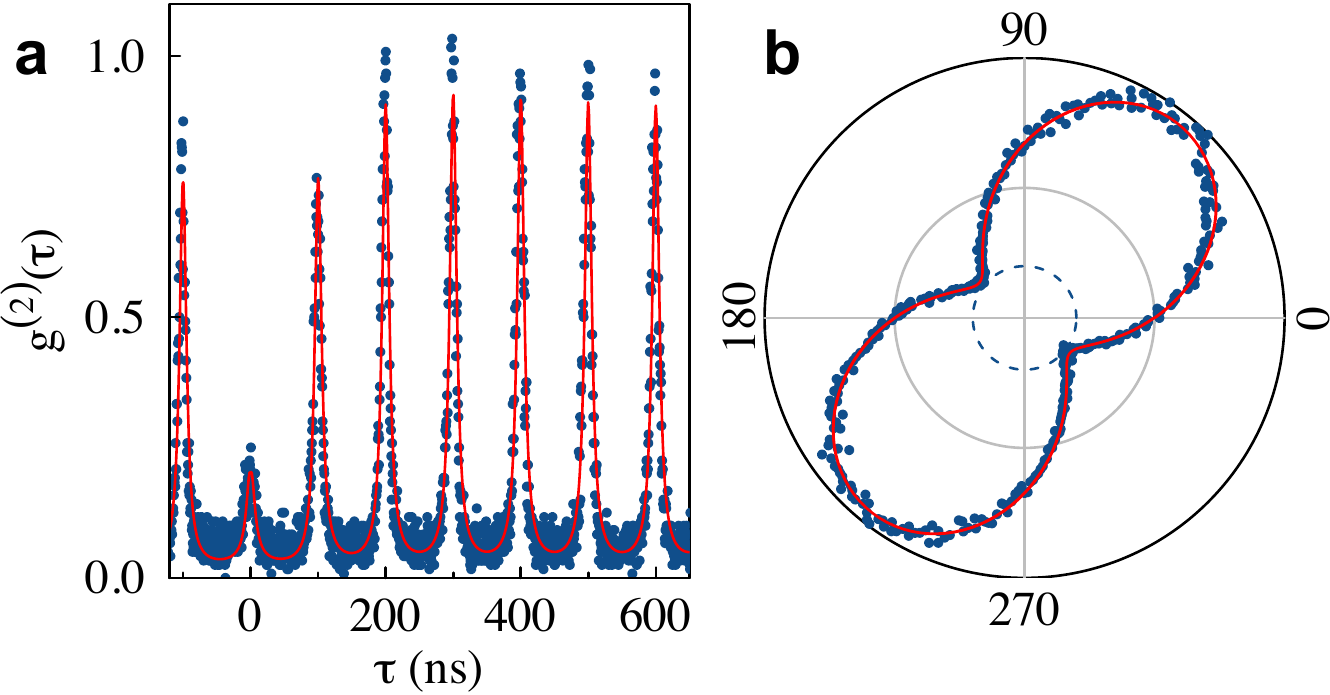}
\end{center}
\caption{ \textbf{Electrically pulsed excitation and polarisation dependence.} \textbf{a}, Second order autocorrelation without background correction showing $g^{(2)}(\tau=0)<0.2$ under 10~MHz pulsed excitation where the pulse width and height were 20~ns and 10~V, respectively. \textbf{b}, Polarisation dependence of an electrically driven single emitter. The dashed circle represents the background level.}
\label{Fig5}
\end{figure}

\noindent {\bf Electrically pulsed single photon emitting diode.} The single photon emitting diodes allow the application of narrow electrical pulses for the production of near-deterministic single photons. A function generator was used to apply 20~ns voltage pulses at 10~MHz in forward bias with a peak voltage of 10~V to inject the charge carriers. A typical pulsed $g^{(2)}(\tau)$ is shown in Fig.~\ref{Fig5}a. The pulsed emission is indicated by the periodic correlation peaks. The height of the zeroth-order peak drops to $g^{(2)}(\tau = 0) \approx 0.2$,  again unambiguously indicating single photon emission and in fair agreement with Fig.~\ref{Fig3}. The non-zero value at $\tau = 0$ can be accounted for by the background EL from the D$_{1}$ center arising from the implantation damage as discussed above. It should be noted that this can presumably be further reduced with the optimisation of the device geometry. The count rate under these conditions was 12.2~kHz giving $\eta_{det.} \eta_{col.} \eta_{def.}=1.2\times 10^{-3}$ where $\eta_{x}$ are the detection, collection and defect photon generation efficiencies in our system, respectively. An estimation of $\eta_{def.}$ is determined to be $>8\%$ with $\eta_{det.}\approx 53\%$ and $\eta_{col.}\approx 3\%$. Since the device is not yet in the saturation regime (Fig.~\ref{Fig4}), this efficiency is more related to an effective recombination efficiency of the defect rather than its intrinsic quantum efficiency. 

Excitation repetition rates greater than 10~MHz decrease the count rate but this may be improved once integrated with a cavity or solid immersion lens \cite{Widmann2014} since $\eta_{col.}$ dominates the total efficiency of the system. To our knowledge, this is the first demonstration of a pulsed room temperature single photon emitting diode.

\noindent {\bf Polarization of single centers.} A $\lambda$-half wave plate and a polarizer were used to rotate and filter the single photon emission. Fig.~\ref{Fig5}b shows the polarisation behavior for a single defect. As indicated by the dashed line, the emission intensity drops to the background level. A study of the polarisation axis of more than 30 defects revealed the same fully polarized emission along three main polarisation axes separated by a 60$^{\circ}$ angle. This implies that the defect emission dipoles are oriented along the hexagonal basal plane of the SiC matrix. The same behavior was observed in PL in absorption and emission. The close similarity of the PL and EL polarisation behavior further indicates that all defects studied here have the same symmetry and therefore suggests the same origin.
\newline

\noindent {\bf\large Discussion}

The CMOS compatibility of SiC lends unprecedented flexibility to further development of this electrically driven single photon source. This includes further optimisation of the device, its integration with optical components such as solid immersion lenses, cavities and on-chip waveguides and finally, device scale-up. Further to this, some flexibility in the integration of other defect types, such as the V$\rm _{Si}$ or the $\rm C_{Si}V_{C}$, is also possible. For instance, defects with a lower thermal stability can be introduced at a later stage in the device processing. Optical filtering would then be needed to target specific defects. 

Further optimisation of the device design is expected to allow higher current densities to be generated at the defect site. This may result in single photon count rates in the MHz regime. The count rate saturation regime (Fig.~\ref{Fig3}a) may then also be accessed to investigate the emission dynamics and quantum efficiencies in more detail. With these developments and the favourable intrinsic properties of these emitters such as their full polarisation (Fig.~\ref{Fig4}b), photo-stability and room temperature operation, opportunities for the application of SiC to quantum information processing, quantum cryptography and single photon metrology are foreseeable. 

\begin{figure}
\begin{center}
\includegraphics[width = 0.5\textwidth]{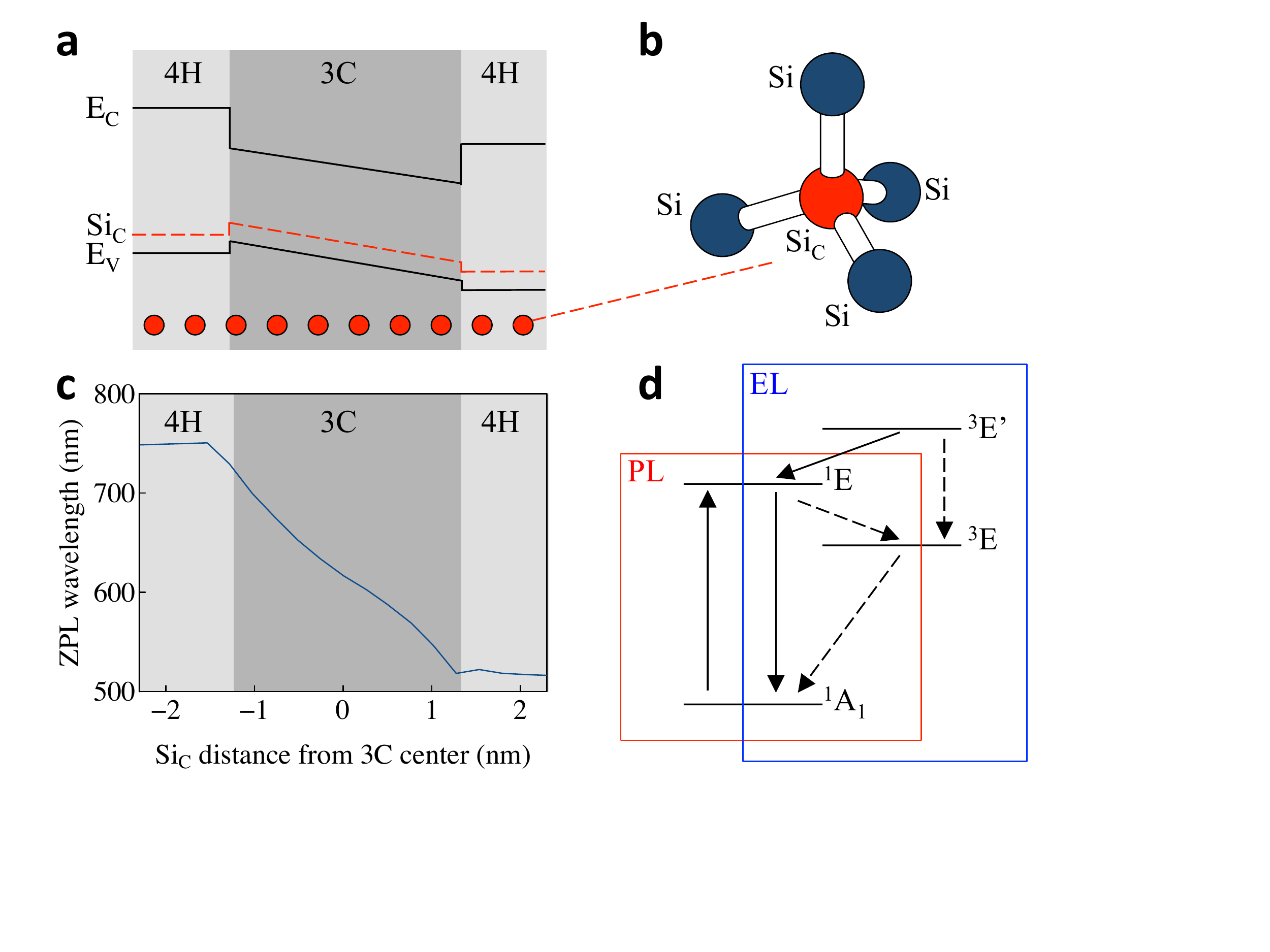}
\end{center}
\caption{ \textbf{Defect origin.} \textbf{a}, Schematic of the band edges of a 3C polytype inclusion in 4H-SiC \cite{Davydov2006}. The x-axis is parallel to the c-axis with increasing depth to the left.  The location of a defect state introduced by the Si$\rm _C$ defect shown in \textbf{b} is symbolized by a red dot where the hole is strongly localized upon excitation. \textbf{c} The calculated ZPL wavelength as a function of the Si$\rm _C$ location around or inside a 2.5~nm wide 3C polytype inclusion along. \textbf{d} A schematic of the energy level diagram of the Si$\rm _C$ defect where the ground state is $\rm ^{1}A_1$. Fast (slow) processes are labeled by straight (dashed) arrows. PL and EL boxes indicate the states participating in the respective processes. The $\rm ^3E'$ state consists of a valence band derived hole captured by the defect whereas the $\rm ^1E$ and $\rm ^3E$ excited states consist of a hole localized on the defect. In these excited states the electron lies on a split conduction band state thus the energies of the $\rm ^3E'$, $\rm ^1E$ and $\rm ^3E$ levels depend on the position of the conduction band edge.}
\label{Fig6}
\end{figure}

As mentioned above, the spectral characteristics of these sources have not previously been reported. We now discuss their possible atomic origin. Their characteristic properties are their high thermal stability, preferential formation close to the surface, wide variability in ZPL and strong polarisation perpendicular to c-axis. These common characteristics suggest a common origin. Firstly, it is known that stacking faults (SFs) often form close to the surface during post-growth surface processing \cite{Brillson2004}. The variation in size and type of these SFs could explain the large ZPL variability \cite{Iwata2003}. However, the polarization of the emitted photons from SFs would be emitted parallel to the c-axis because of the spatial separation of the electron and hole along the c-axis in the excited state \cite{Brillson2004}. This is in contrast to the observations made here. Alternatively, the polarisation and high thermal stability may be explained by a stable point defect with a high $C_{3v}$ symmetry. However, a single defect type residing on inequivalent sites cannot account for the observed 200~nm variance in ZPL position. We therefore conclude that the most likely origin is a combination of these two structures, namely, a thermally stable $C_{3v}$-symmetry point defect which is near or embedded into a stacking fault where the luminescence of the defect involves states split from the conduction band edge as shown in Fig.~\ref{Fig6}a. Such a model accounts for the spatial location of these sources, their polarisation and the fact that their density cannot be enhanced by electron irradiation since the SF concentration is intrinsically low. Very few defects show a thermal stability up to 1600$^\circ$C in 4H-SiC. One such defect is the $D_1$ center (Si$\rm _C$ shown in Fig.~\ref{Fig6}b). In our devices, luminescence from this center can be clearly observed (for example in Fig.~\ref{Fig1}d) and therefore certainly has high thermal stability. The D$\rm _1$ center also has a high $C_{3v}$ symmetry with polarization perpendicular to the c-axis \cite{Egilsson1999}. The luminescence mechanism can be described as recombination of the electron at the split conduction band state and a localized hole on the defect site.

Ab-initio calculations (see Methods) of a Si$_{\text{C}}$ defect in proximity to a single cubic SF show that the ZPL energy is indeed sensitive to the defect-SF distance while also maintaining the same general recombination behavior found for the defect in bulk 4H-SiC. Fig.~\ref{Fig6}c shows that when the defect is placed close to the SF, the ZPL energy follows the variation in the conduction band edge induced by the SF. As the defect-SF distance increases the value tends to the bulk value. We considered a 2.5-nm wide 3C polytype inclusion. This structure yields a triangular potential well for the conduction band electrons. We find that defects close to but above the 3C inclusion along the c-axis emit light around 540~nm. Defects located inside the 3C inclusion can emit a continuous range of longer wavelength photons from the top side down to the bottom side of the inclusion. Defects beneath the 3C inclusion show emission around 750~nm. A variation in defect position near or inside this 3C inclusion could then explain the broad variance in ZPL position and give rise to the histogram displayed in Fig.~\ref{Fig2}d. 

These calculations also allowed the electronic structure and recombination kinetics at the Si$_{\text{C}}$ defect to be considered in detail (Fig.~\ref{Fig6}d). The highest fully occupied defect level has an $e$ character in the ground state which lies at about 0.3~eV above the valence band edge in 4H-SiC, and no empty defect state appears in the gap. In the excited state the electron occupies an $a_1$ state split from the conduction band edge and leaves a hole in the strongly localized $e$ state. This can form an optically active $^1E$ excited state and a lower-energy $^3E$ shelving state. The strong luminescence between $^1E$ and $^1A_1$ only produces photons with a polarisation perpendicular to the c-axis. Interestingly, the presence of this $^3E$ state could explain the photo-dynamics we observe in the PL. We also found that the $e$ state in the excited state shifts down toward the valence band edge compared to the case of the ground state. In the EL process the hole may first be captured by the valence band derived $e$ state forming a higher-energy $^3E^\prime$ state that can decay toward the optically active $^1E$ excited state by spin-orbit interaction mediated by phonons. This may explain the four-level-like EL process described in Fig.~\ref{Fig3}. Our working model can basically account for all properties of the PL and EL single photon sources observed in this work. Nevertheless, we do not discount other possible candidate extrinsic or intrinsic defects that have a similar electronic structure. 

% Our ab-initio calculations show that the highest fully occupied defect level has an $e$ character in the ground state which lies at about 0.3~eV above the valence band edge in 4H-SiC, and no empty defect state appears in the gap. 
% This state has an $^1A_1$ character. ???  
  
In conclusion, electrically driven single photon sources operating at room temperature have been integrated into a range of SiC devices. This is achieved using CMOS compatible fabrication techniques, namely photolithography, dopant implantation and annealing. The high temperature processing resulted in highly efficient single photon emitters with a stable and high emission rate that exceed that of other room temperature electrically driven single photon sources. These defect-based sources have a high thermal stability, wide variability in ZPL wavelength and a strong polarisation. The recombination kinetics EL mode showed characteristics of a double transition before radiative decay while PL model could be characterised by a standard three level energy level scheme.  A model is proposed which is able to account for all of these traits. It consists of a $Si_C$ in close proximity to a SF-type defect. Ab-initio calculations are performed to support the validity of this model. The realization of bright stable electrically driven single photon emission in SiC, in addition to the demonstration of pulsed excitation of the fully polarized single photons, is an important step towards on-demand single photons for applications in quantum information processing, quantum cryptography and single photon metrology. 
\newline

\noindent {\bf\large Methods}
{\small 

\noindent {\bf Device fabrication.} The devices were fabricated on epilayers grown on 4H-SiC substrates by chemical vapor deposition purchased from CREE. The thickness and doping concentration was 4.9~$\mu m$ and $4\times 10^{15} \;\rm cm^{-3}$, respectively. Multiple energy Al implantation was performed into the epilayers while the sample was held at 800$^{\circ}$C to form the p$^{+}$ regions. The 6H samples are discussed in the Supplementary Information. The implantation energies were chosen so as to form a nearly uniform dopant concentration depth profile over a 150~nm range. Following implantation an anneal of 1800$^{\circ}$C for 10 minutes in an Ar ambient was used to activate the implanted dopants, repair the implantation damage and also form the single photon emitters. The carbon film was removed using a dry oxidation process at 800$^{\circ}$C following the method described in \onlinecite{Negoro2004}. 

A 1100$^{\circ}$C pyrogenic oxidation (H$_2$:O$_2$ = 1:1) was then performed to grow a field oxide. This had the effect of bring the SiC surface closer to the peak in the implanted dopant profile to allow the formation of high quality Ohmic contacts. This oxidation step was also found to reduce the PL background. The contacts were formed by Al deposition and subsequent sintering at 850$^{\circ}$C for 5 minutes. A back contact was formed by Al deposition without sintering. Full electrical characterisation was then performed using current-voltage and capacitance-voltage measurements. The PL experiments were performed with HPSI 4H-SiC from CREE. Selected samples were irradiated with 2~MeV electrons to a fluence of $1\times 10^{17}\rm\; cm^{-2}$ while the sample temperature was kept below 80$^{\circ}$C in a N$_{2}$ ambient to create a homogeneous defect density. Both as-grown and irradiated samples were annealed at 1650$^{\circ}$C for 10 minutes in Ar. Carbonisation of the surface and a subsequent 1100$^{\circ}$C oxidation of these samples were also investigated. 

\noindent {\bf Confocal fluorescence microscopy.}
A custom confocal system consisting of a HBT interferometer and single-photon-sensitive avalanche photodiodes was used to measure the time correlation of the fabricated single photon emitters. Photon counting and correlation were carried out using a time-correlated single-photon-counting module (PicoHarp 300, PicoQuant GmbH). The devices were driven in forward bias and the EL collected with a $\times$100 infinity-corrected air objective lens with a NA of 0.95. The luminescence was collected through a 50~$\mu$m multi-mode fibre pinhole. Optical excitation was achieved with a 532~nm continuous-wave laser directed through the objective via a dichroic mirror. Spectra were collected with a spectrometer (Princeton Instruments Acton 2300i, Acton) with a Peltier-cooled CCD. Samples were mounted on a piezoelectric XYZ stage allowing 100 $\times$ 100 $\rm\mu m ^{2}$ scans. The laser line or, during EL experiments, the D$_{1}$ line was spectrally removed with a 560-nm long-pass filter. The single-photon measurements were performed at room temperature.
For low temperature measurements the sample was mounted in nitrogen cooled cryostat (Linkam THMS600). Instead of a XYZ stage, a piezoelectric scanning mirror was used. The emission was excited and collected through a cover slip corrected 40x air objective with a NA of 0.6. Optical excitation was driven by either a 532~nm or a 633~nm continuous-wave laser. The 633~nm laser was required to excite emitters with ZPL positions at higher wavelengths.

\noindent {\bf Modelling.}
The atomistic simulations were carried out using density functional plane wave supercell calculations. Our working model is the isolated silicon antisite defect (Si$_{\text{C}}$) in bulk and near or inside regularly ordered stacking faults, i.e., polytype inclusions. We studied the Si$_{\text{C}}$ defect at the hexagonal site of 4H-SiC in detail. We do not find qualitative differences in the ground state properties of Si$_{\text{C}}$ defect at the cubic or hexagonal sites. The Si$_{\text{C}}$ defect has a $C_{3v}$ symmetry. In order to eliminate the finite size effect of the simulated cell we applied a very large supercell consisting of 1536 atoms of bulk 4H-SiC in which to embed the defect. We applied $\Gamma$-point sampling of the Brillouin-zone that is sufficient to provide convergent charge density and enabled us to study the degeneracy of the defect levels. We applied the Perdew-Burke-Ernzerhof (PBE) semilocal DFT \cite{PBE} in most of the calculations. The PBE functional introduces a band gap error of about 0.9~eV in SiC. However, we checked the ground state properties of the Si$_{\text{C}}$ defect in a 576-atom supercell with HSE06 hybrid DFT \cite{HSE06} which is able to reproduce the experimental band gap of 4H-SiC ($\approx$~3.2~eV) and defect levels in the fundamental band gap \cite{Deak10, Ivady11}. By comparing the PBE and HSE06 results we found that a rigid scissor correction on the conduction band edge in the PBE calculation provided a very good description of the single particle levels of the defect state and conduction band edge with respect to the calculated valence band edge. The value of this scissor correction for the zero-phonon line (ZPL) peak of Si$_{\text{C}}$ defect is 0.81~eV.\\
The zero-phonon line energies were calculated with the constraint DFT (CDFT) method where the excited state is created by promoting a single electron from the occupied state to the empty state \cite{Gali09}. We also calculated ab-initio the ZPL peak of the Si$_{\text{C}}$ defect near one cubic stacking fault in 4H-SiC that may be considered as a half unit of a 6H polytype inclusion. As a consequence, the band gap of the system is reduced by 0.23~eV due to the new conduction bands introduced by the stacking fault acting as a quantum well for electrons. We used a 3016-atom supercell for this study which involves five 4H-SiC unit cells and two extra cubic bilayers (stacking faults) along the c-direction of the lattice. Two stacking faults are needed because of the periodic boundary condition along the c-axis. These two stacking faults were placed as far away from each other as possible ($\sim$2.5~nm). We placed the Si$_{\text{C}}$ defect at a hexagonal site nearest to the stacking fault and then three and five bilayers apart. We found that the calculated ZPL peaks of these defects approach the ZPL peak of Si$_{\text{C}}$ defect in perfect 4H-SiC as the Si$_{\text{C}}$ defects lie farther from the stacking fault. The same phenomena is expected to arise for larger polytype inclusions such as 3C inclusions that should contain a minimum of three consecutive cubic bilayers. However, these structures are not computationally feasible at the ab-initio level. \\
For realistically large 3C polytype inclusions in 4H-SiC we applied a simplified quasi one-dimensional model which adequately describes the electronic structure of the defect either near or inside the 3C polytype inclusions (see \onlinecite{Tetrapods2014} for technical and other details). The code used to calculate the ZPL peaks of polytype inclusions \cite{Tetrapods2014} was modified so that the hole is localized at the defect site whereas the electron occupies the lowest-energy empty state in the excited state of the defect. The localization of the hole is achieved by adding an attractive potential for the hole. This attractive potential was set to reproduce the experimental ZPL peak of the Si$_{\text{C}}$ defect in bulk 4H-SiC and then the same attractive potential was applied in the polytype inclusion model. This methodology was justified from ab-initio calculations on the Si$_{\text{C}}$ defect in bulk 4H-SiC and near to a stacking fault and showed that the hole is a deep defect state localized around the defect where its level lies at about 0.3~eV above the valence band edge.\\

{\small
\noindent {\bf Acknowledgements.}\\
 B.C.J. acknowledges the Australian Research Council center for Quantum Computation and Communication Technology (CE110001027) and the Dyason Fellowship for financial support. T.O. acknowledges the Ministry of Education, Science, Sports and Culture, Grant-in-Aid (B) 26286047. A.G. acknowledges the Lend\"{u}let program of the Hungarian Academy of Sciences.\\

\noindent {\bf Author contributions.} \\
 A.L. and B.C.J. designed and performed the experiment, did the data analysis and prepared the manuscript. A.G. and Z.B. performed the ab initio calculations and related analysis. N.I. and T.O. fabricated the devices. All authors discussed the results and contributed to the manuscript. \\

\noindent {\bf Additional information}\\
\noindent{\bf Supplementary Information} accompanies this paper.
 
\noindent{\bf Competing Interests:} The authors declare that they have no competing financial interests.
 
\noindent{\bf Correspondence:} 
 Correspondence and requests for materials should be addressed to B.C.J. \\(email: johnsonb@unimelb.edu.au).
}

% \bibliography{library1}


\end{document}

% \title{Electrically driven single photon source at room temperature}
\title{Supplementary Information \\ Single-photon emitting diode in silicon carbide}
% Triggered single photon emitting diode

\author{A. Lohrmann}
\affiliation{%
School of Physics, The University of Melbourne, Victoria 3010, Australia
}%

\author{N. Iwamoto}
\affiliation{%
SemiConductor Analysis and Radiation Effects Group, Japan Atomic Energy Agency, 1233 Watanuki, Takasaki, Gunma 370-1292, Japan.
}%

\author{Z. Bodrog}
\affiliation{Institute for Solid State Physics and Optics, Wigner Research Centre for Physics, Hungarian Academy of Sciences, Budapest, POB 49, H-1525, Hungary} 

\author{S. Castelletto}
\affiliation{%
School of Aerospace, Mechanical and Manufacturing Engineering RMIT University, Melbourne, Victoria 3001, Australia RMIT University.
}%

\author{T. Ohshima}
\affiliation{%
SemiConductor Analysis and Radiation Effects Group, Japan Atomic Energy Agency, 1233 Watanuki, Takasaki, Gunma 370-1292, Japan.
}%

\author{T. J. Karle}
\affiliation{%
School of Physics, The University of Melbourne, Victoria 3010, Australia
}%

\author{A. Gali}
\affiliation{Institute for Solid State Physics and Optics, Wigner Research Centre for Physics, Hungarian Academy of Sciences, Budapest, POB 49, H-1525, Hungary} 

\affiliation{%
Department of Atomic Physics, Budapest University of Technology and Economics, Budafoki \'{u}t 8, H-1111, Budapest, Hungary 
}%

\author{S. Prawer}
\affiliation{%
School of Physics, The University of Melbourne, Victoria 3010, Australia
}%

\author{J. C. McCallum}
\affiliation{%
School of Physics, The University of Melbourne, Victoria 3010, Australia
}%

\author{B. C. Johnson}
\affiliation{%
Centre for Quantum Computing and Communication Technology, School of Physics, University of Melbourne, Victoria 3010, Australia.
}%

\maketitle

\section{Material characteristics}

% EL occurs above an onset voltage of V$_\text{on} = 2.7 \text{V}$ 

\begin{figure}
\includegraphics[height=9cm]{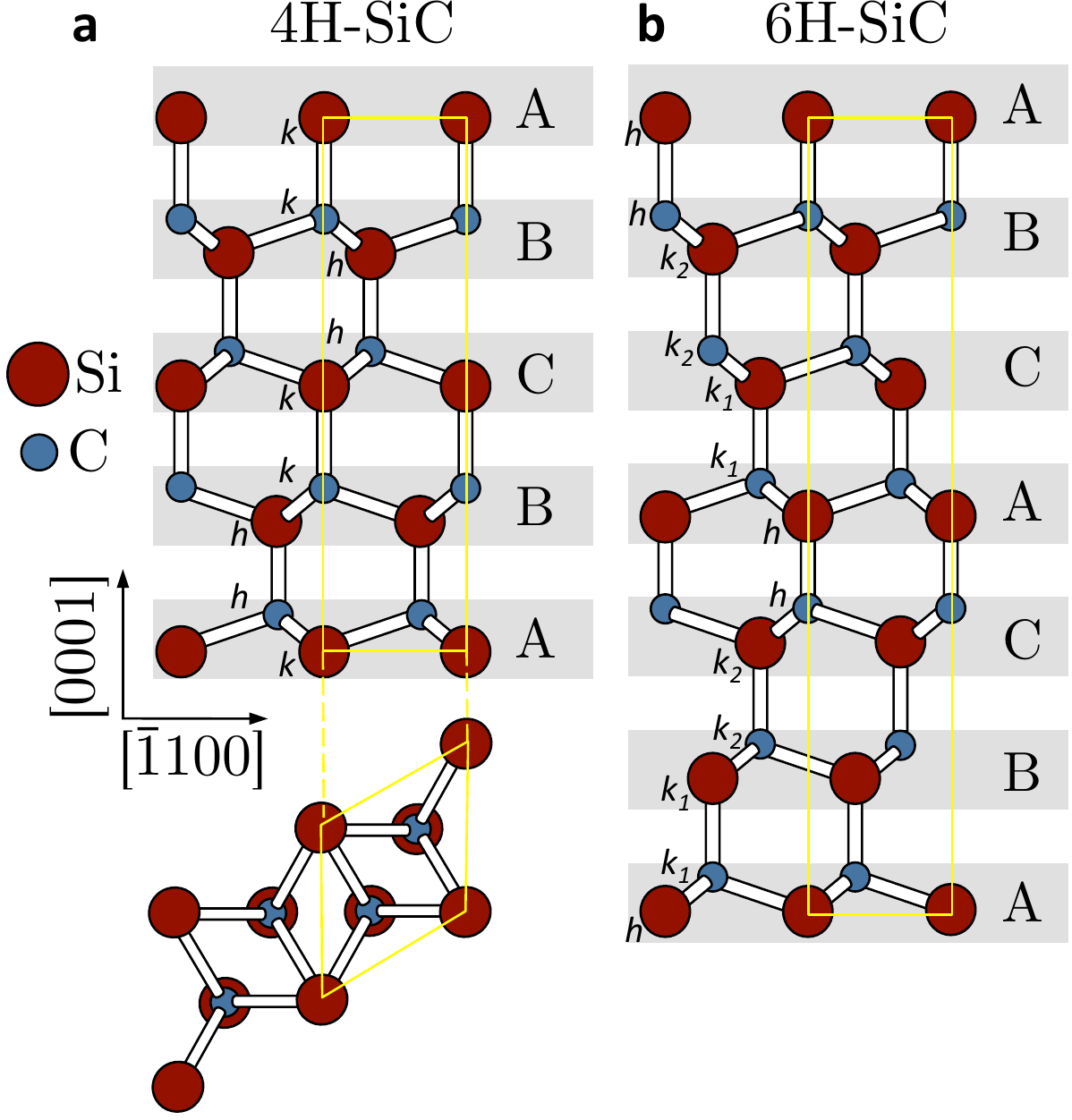}
\caption{{\bf Atomic structure of hexagonal SiC.} Two unit cells of {\bf a}, 4H-SiC and {\bf b}, 6H-SiC projected on the (11$\bar 2$0) plane with lengths of 10.048~{\AA} and 15.1248~{\AA}, respectively. A projection along $<$0001$>$ is also shown under the 4H-SiC to illustrate the hexagonal configuration of atoms along the basal plane. Kinks along the stacking direction, [0001] arising from a vector translation of the bilayers (shaded regions) periodically appear and differentiate the polytypes. Point defects of the same type may have different spectral properties depending on whether they occupy an inequivalent hexagonal or cubic lattice site denoted by the $h$ and $k$ of which there are two in 4H and three in 6H SiC.}
\label{S1}
\end{figure} 

Electrically driven single photon sources operating at room temperature have also been integrated into both 4H and 6H SiC devices. These two polytypes are of great technological importance and can be grown as virtually dislocation-free 3 inch wafers. The basic structure is a covalently $sp^{3}$ bonded network of alternating Si and C atoms. The various polytypes arise from the different ways the SiC bi-layers are stacked along the c-axis as shown in Fig.~\ref{S1}. 

\begin{figure}[htbp]
\includegraphics[width=12cm]{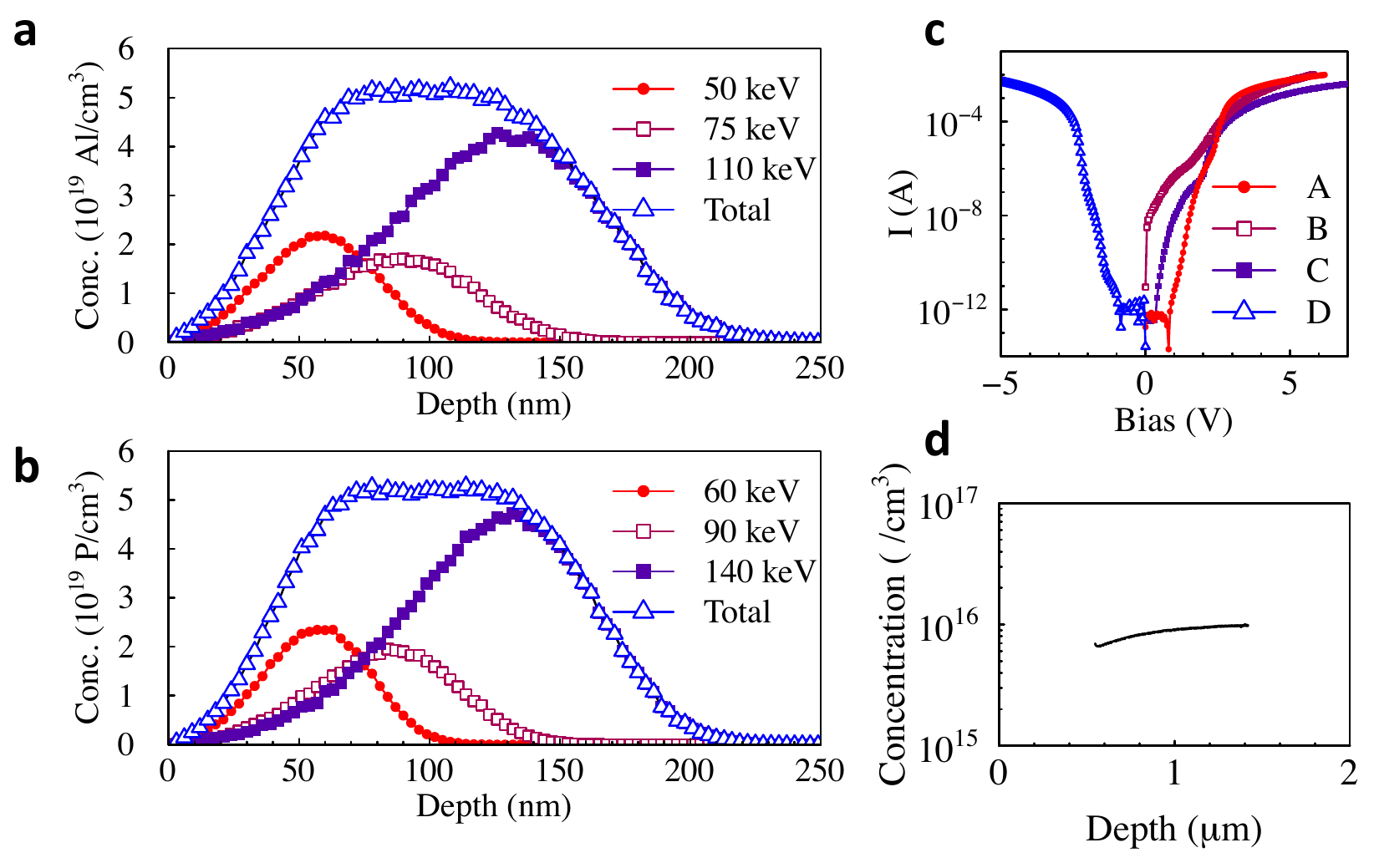}
\caption{{\bf Device fabrication and characteristics} Monte Carlo simulations of {\bf a}, the Al implantation profile and {\bf b}, the P implantation profile after multi-energy implantations. {\bf c}, The typical forward bias current-voltage characteristics for a SiC diodes containing single photon sources. Samples A-D are 6H p$^{+}$n,  6H p$^{+}$n,  4H p$^{+}$n,  and 6H n$^{+}$p,  respectively. {\bf d}, The carrier concentration versus depletion width in the epi-layer of sample C determined with capacitance-voltage measurements to be $9.8\times 10^{15} \rm\; N/cm^{3}$.}
\label{S2}
\end{figure} 

Several batches of devices composed of both 4H and 6H SiC were fabricated using the procedure outlined in the methods section of the main text. To predict the profile of implanted species used to form the p$^{+}$n or n$^{+}$p junctions we used a Monte Carlo simulation (SRIM code) \cite{SRIM}, the Al and P results are shown in Fig.~\ref{S2}a and \ref{S2}b, respectively. Multiple energies were used in both cases to form a broad constant concentration profile with a peak concentration of $5\times 10^{19}\;\rm cm^{-3}$. Al implants of 110, 75 and 50~keV were performed with fluences of $4.2\times 10^{14}$, $1.3\times 10^{14}$ and $1.2\times 10^{14} \rm\; Al/cm^{2}$. The energies and fluences of the P implants were 140, 90 and 60 keV and $4.2\times 10^{14}$, $1.3\times 10^{14}$ and $1.2\times 10^{14}\;\rm P/cm^{2}$, respectively. During implantation the samples were held at 800$^{\circ}$C to minimise defect formation and avoid amorphisation. 

% One of the HPSI samples purchased from CREE was irradiated with 2~MeV electrons with a fluence of $1\times 10^{17}$ per cm$^2$. The energy a homogeneous distribution of defects throughout the sample. The defect density, however, as measured with the confocal microscope did not significantly increase for this sample.

Figure~\ref{S2}c and \ref{S2}d show the forward current characteristics of four different devices with a 1~mm diameter top contact and the carrier concentration determined in reverse bias from capacitance measurements, respectively. As expected for such SiC diodes, the reverse current of all devices tested remained in the sub-nA range for voltages with magnitudes up to 200~V. 

%%%%%%%%%%%%%%%%%%%%%%%%%%%%
%%%%%%%%%%%%%%%%%%%%%%%%%%%%
%%%%%%%%%%%%%%%%%%%%%%%%%%%%

\section{Electrically driven single photon sources in 6H SiC}

\begin{figure*}[htbp]
\begin{center}
\includegraphics[width=12cm]{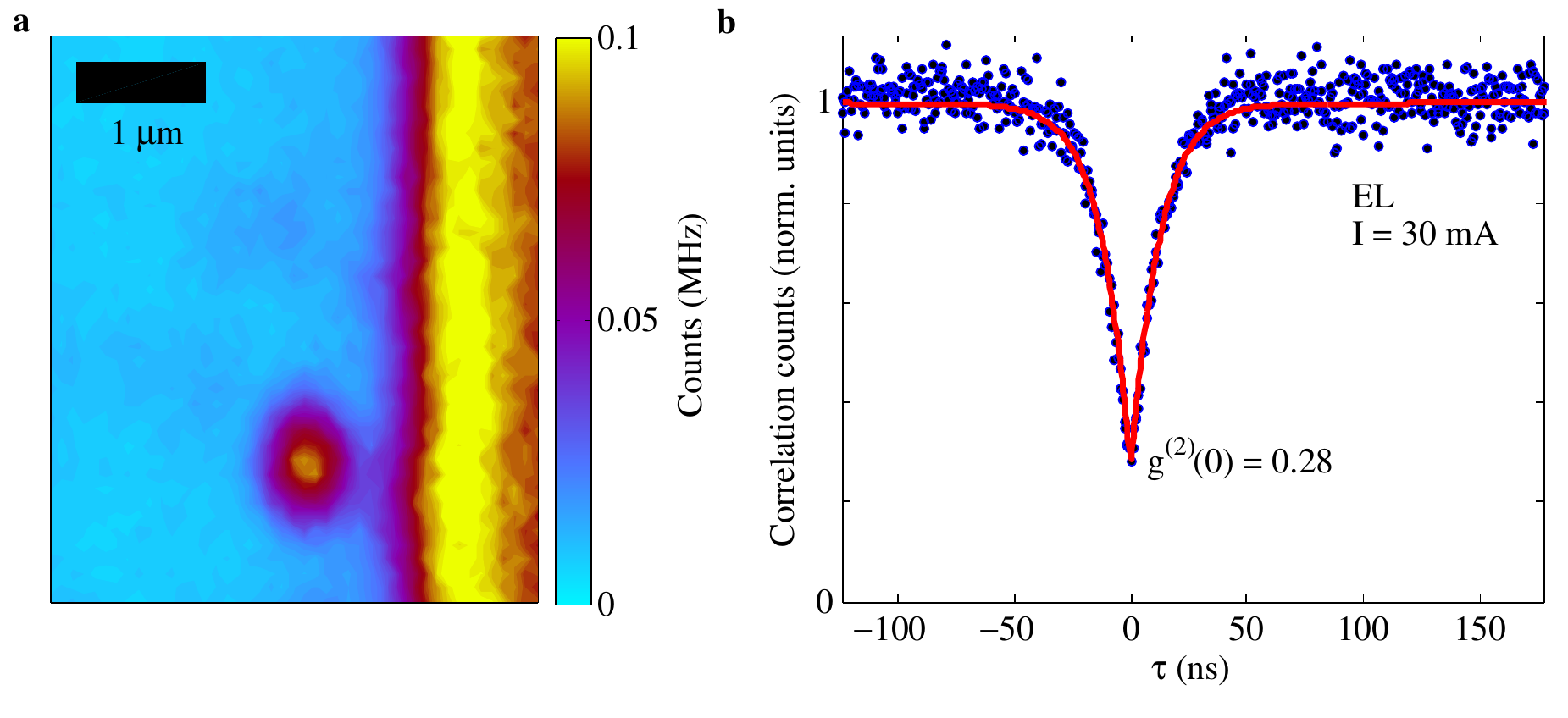}
\end{center}
\caption{\textbf{Electrically driven single defects in 6H-SiC. a}, Confocal map of a single defect in a 6H SiC device. The single defect resides between the floating guard rings which display strong D$_{1}$ emission in this device. {\bf b}, Autocorrelation measurement of the single defect in {\bf a}. }
\label{S3}
\end{figure*}

\begin{figure*}[hbp]
\begin{center}
\includegraphics[width=14cm]{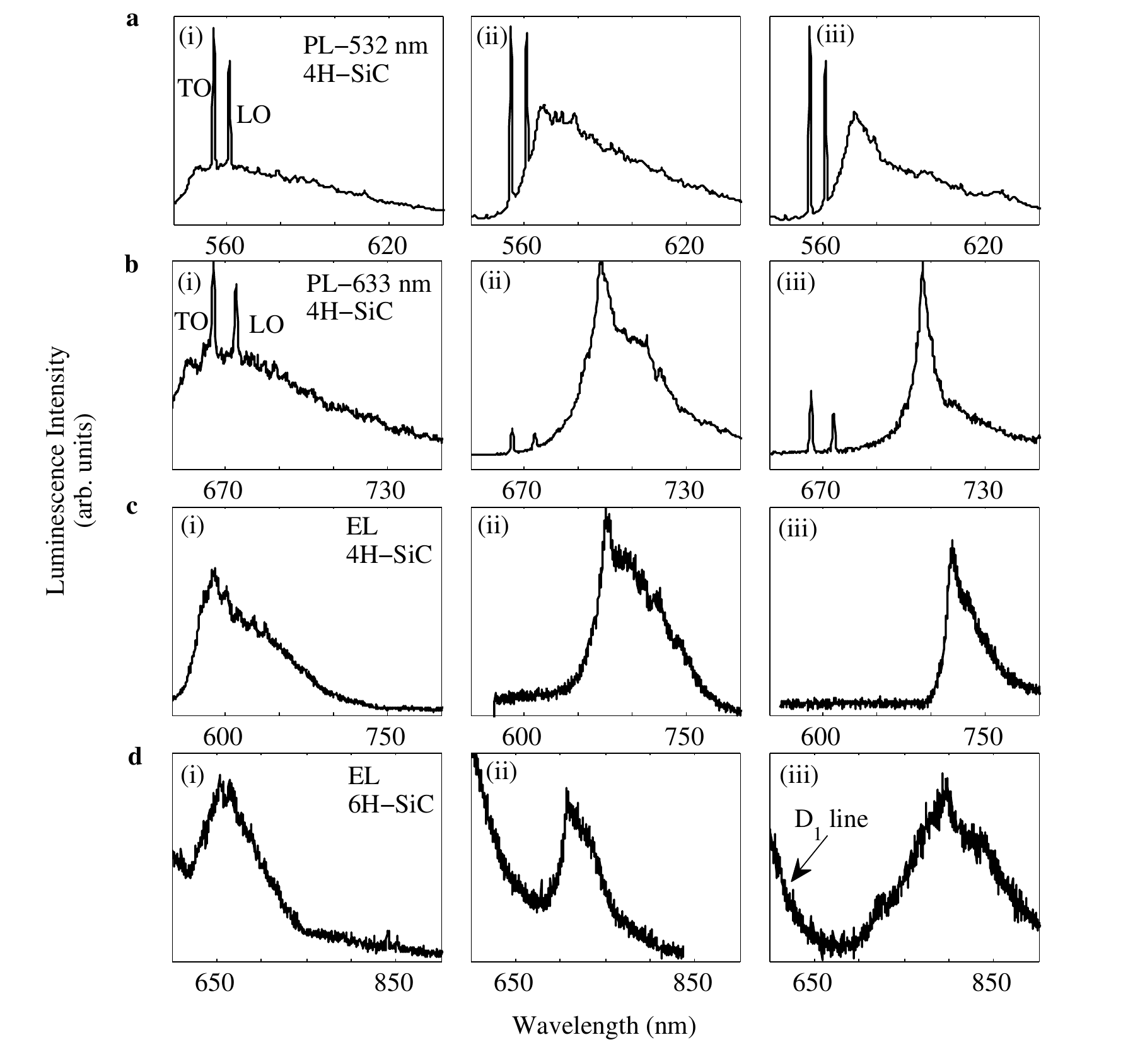}
\end{center}
\caption{{\bf Spectral characteristics of single photon emitters in 4H-SiC.} {\bf a}, Several different but reproducible PL spectra excited with {\bf a, }532~nm and {\bf a, } 633~nm. EL spectra for single defects in {\bf c, } 4H-SiC and {\bf d, } 6H-SiC.}
\label{S4}
\end{figure*}

As mentioned in the main text, single photon emitters were also integrated into 6H-SiC devices. The properties of these emitters were very similar to those found in the 4H-SiC. A typical EL confocal map of a single defect in 6H-SiC is shown in Fig.~\ref{S3}a. This particular device is a P implanted n$^{+}$p junction diode. We observed that the floating guard rings were bright with D$_{1}$ emission and this is evident as a bright band next to the single defect. The D$_{1}$ center's PSB exhibits a 30~nm red shift with respect to the 4H-SiC devices and so, with a similar filter to that used for the 4H-SiC measurements, can result in a poorer signal to noise ratio. The high recombination rate at the D$_{1}$ center in this device may also explain the general lower emission rates of the single photon emitters. 

%The $g^{(2)}(\tau)$ measurement of this center is shown in Fig.~\ref{S4}b and proves it to be a single defect. Again the characteristic plateau near $\tau=0$, which is similar to that in the 4H-SiC devices. An excellent fit is achieved with the four level model which is discussed in detail in the following section. The dynamics of this defect are similar to the defect presented in the main text, with an excited state lifetime of 5.1(2.0)~ns.

The $g^{(2)}(\tau)$ measurement of this center is shown in Fig.~\ref{S3}b and proves that it is a single defect. The dynamics of defects in 6H-SiC are similar to 4H-SiC, which are presented in the main text. This particular defect does not exhibit strong photon bunching, however, the count rates are still low compared to the 4H emitters.

%Spectral analysis revealed several different defect types in the spectral region from 600 to 850~nm.  The spectra were dominated by the very bright D$_{1}$ center emission, which was removed from the spectra via post-processing in order to highlight the other spectral features. The residual single defect spectra for six different defect types are shown in Figure \ref{S4}c. The spectra are shifted with respect to the 4H defects towards longer wavelengths which can be attributed to the same mechanism that causes the D$_{1}$ centre emission to shift. As with the 4H devices, PL measurements were not possible due to the high doping concentration.

Spectral analysis revealed several different defect types in the spectral region from 600 to 850~nm.  The spectra were dominated by the very bright D$_{1}$ center emission. Single defect spectra for three different defect types are shown in Figure \ref{S4}d. The spectra are shifted with respect to the 4H defects towards longer wavelengths which can be attributed to the same mechanism that causes the D$_{1}$ centre emission to shift. As with the 4H devices, PL measurements were not possible due to the high doping concentration.
%%%%%%%%%%%%%%%%%%%%%%%%%%%%
%%%%%%%%%%%%%%%%%%%%%%%%%%%%
%%%%%%%%%%%%%%%%%%%%%%%%%%%%

\section{EL and PL spectroscopy of single defects}

%A wide range of spectra associated with both electrically and optically driven single photon sources is found. A selection of single defect spectra is shown in Fig.~\ref{SI4}. A survey of 100 defects revealed that many spectral features are located in the 600-700~nm range and are dominated by a strong PSB.

A wide range of spectra associated with both electrically and optically driven single photon sources is found. A selection of single defect spectra is shown in Fig.~\ref{S4}. A survey of 100 defects revealed that many spectral features are located in the 600-700~nm range and are dominated by a strong PSB. Fig.~\ref{S4}a and Fig.~\ref{S4}b show PL spectra obtained from single defects in 4H-SiC in annealed HPSI SiC under excitation with 532~nm and 633~nm light. The Raman lines are labelled TO and LO indicating Raman scattering with the transversal and longitudinal optical phonons. The emission spans from the filter edge up to more than 700~nm. The D$_1$ line is not exited by the laser. Additional EL spectra for 4H-SiC and three selected spectra for single defects in 6H-SiC are presented in Fig.~\ref{S4}c and Fig.~\ref{S4}d, respectively.

%%%%%%%%%%%%%%%%%%%%%%%%%%%
%%%%%%%%%%%%%%%%%%%%%%%%%%%
%%%%%%%%%%%%%%%%%%%%%%%%%%%

\section{Polarization in PL}
\begin{figure*}[htbp]
\begin{center}
\includegraphics[width=12cm]{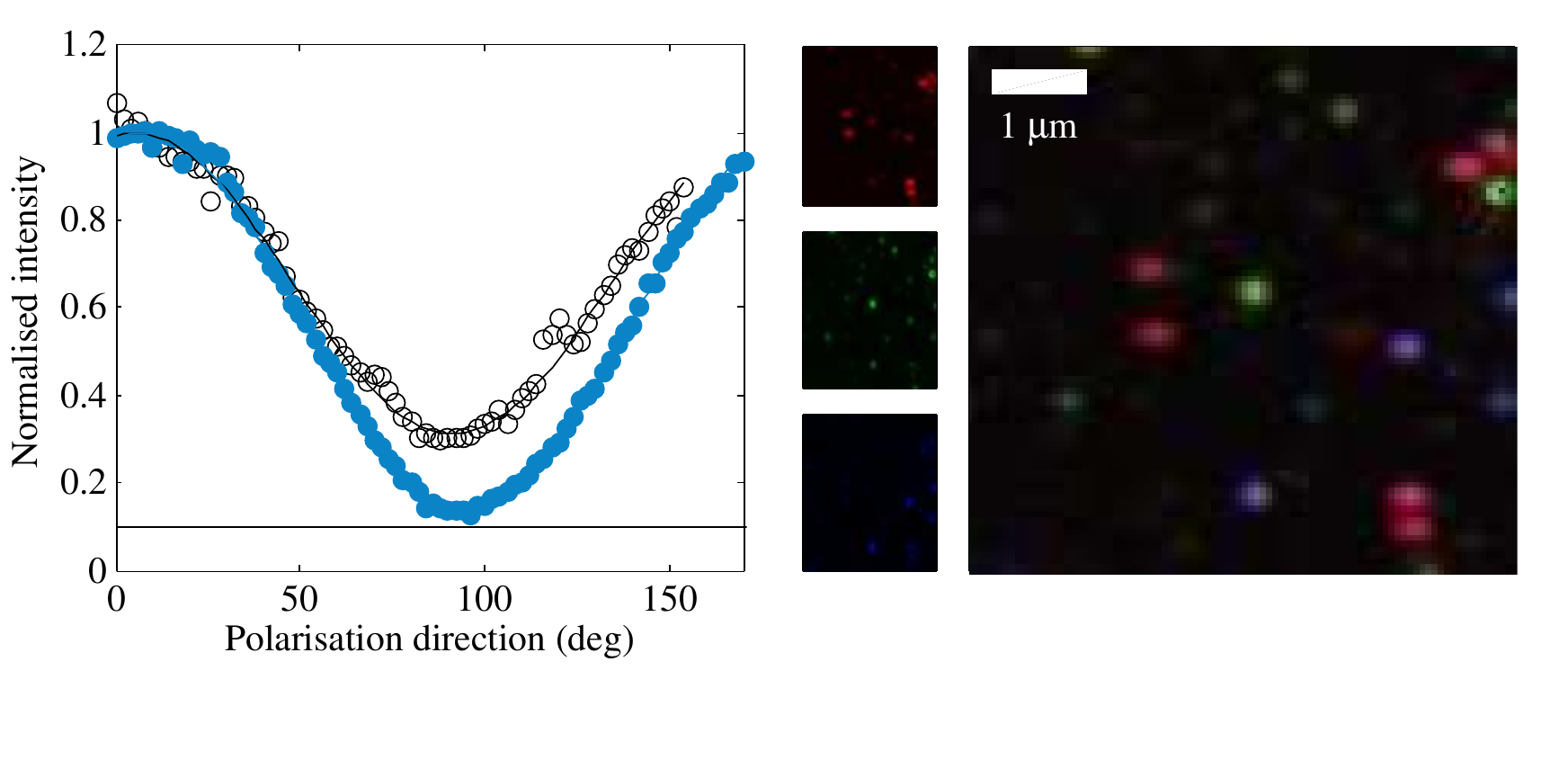}
\end{center}
\caption{{\bf Polarisation behavior. a,} Normalized fluorescence intensity under variation of absorption (black) and emission (blue) polarisation filtering for a single defect. A squared sinusoidal curve is fitted and the dipole orientation is extracted. \textbf{b}, Left: confocal images under three different, relative laser-polarization orientations (red: 30~$^\circ$, green: 90~$^\circ$, blue: 150~$^\circ$). Right: merged image.}
\label{S5}
\end{figure*}

To determine the PL dipole orientation in absorption and emission, PL polarization measurements were performed. Figure~\ref{S5}a shows the fluorescence intensity of a single emitter under varying excitation polarization (black circles) and its emission polarization under fixed excitation orientation(blue circles) relative to each other. The squared sinusoidal fit shows a perfect alignment between the emission and the absorption dipole. A useful measure is the polarization visibility $V = (I_{max}-I_{min})/(I_{max}+I_{min})$, where $I$ is the detector count rate. After subtracting the Raman related background (black line), a visibility of $V_{em}=0.91$ was observed for the emission polarization. The absorption visibility is significantly lower with $V_{abs}=0.59$. This behavior was verified for most of the investigated emitters. The emission visibility is comparable to that measured for EL emitters.\\
Figure~\ref{S5}b shows a set of three confocal maps taken under excitation with three different polarization orientations 60 degrees apart (left). By merging the maps it can be shown that the emitters are polarized along three different polarization axes that are 60 degrees apart.

\begin{figure*}[htbp]
\begin{center}
\includegraphics[width=12cm]{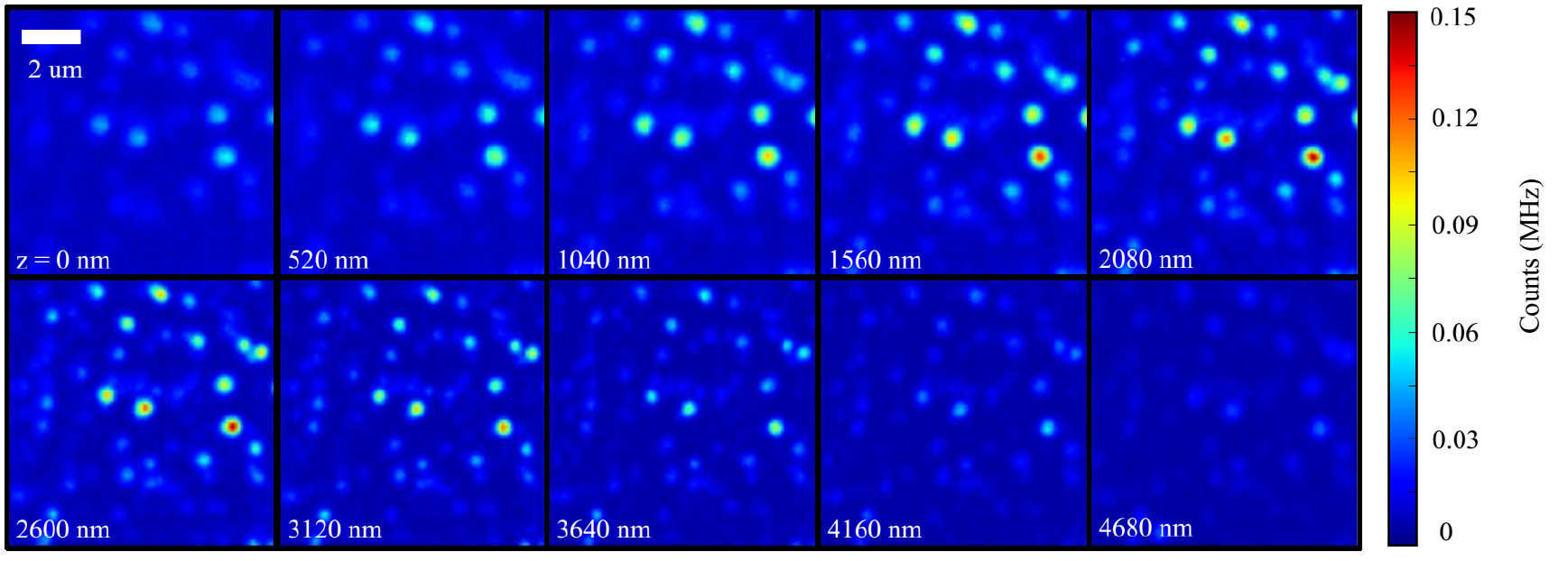}
\end{center}
\caption{{\bf Confocal $x-y$ scan of single photon sources in 4H-SiC as a function of depth.} The confocal maps are collected with 532~nm excitation approximately from the surface to a depth of almost 5~$\mu$m. All centers appear in a narrow band close to the surface.}
\label{S6}
\end{figure*}

\section{Defect location}

The vertical extent of single photon emitters was determined with a series of vertical $x-y$ confocal PL maps taken every 10~nm in depth, $z$, using optical excitation of the HPSI 4H-SiC sample after a 1650$^{\circ}$C anneal followed by a 1100$^{\circ}$C oxidation. Fig.~\ref{S3} shows a selection of these maps. The superior diffraction limited axial resolution of the PL confocal mode over that of EL allowed the approximate depth of the emitters to be determined with a high accuracy. In addition, the apparent distribution of defects observed with EL is a convolution of the actual defect density with the current paths within the device. In Fig.~\ref{S6} it can be seen that the defects exist in a narrow depth band as their intensity increases and decreases over the range of the map series. When the peak intensity of several single emitters versus the depth is fit with a Gaussian distribution after corrected for the sample tilt, the emitters have a mean distance from the tilt plane of just $\approx$40~nm. This position was common throughout all samples studied and is expected to also be true of the pn junction diodes which showed a similar density of alike defects. Therefore, the emitters exist in a narrow band close to the surface.

%%%%%%%%%%%%%%%%%%%%%%%%%%%%
%%%%%%%%%%%%%%%%%%%%%%%%%%%%
%%%%%%%%%%%%%%%%%%%%%%%%%%%%

%removed

%%%%%%%%%%%%%%%%%%%%%%%%%%%%
%%%%%%%%%%%%%%%%%%%%%%%%%%%%
%%%%%%%%%%%%%%%%%%%%%%%%%%%%

%removed

\bibliography{library1}